\documentclass[a4paper,clock]{article}
\usepackage[dvips]{epsfig}
\usepackage{amssymb}
\usepackage{bm}
\usepackage{dcolumn}

\catcode`\@=11
\def\marginnote#1{}
\newcount\hour
\newcount\minute
\newtoks\amorpm
\hour=\time\divide\hour by60
\minute=\time{\multiply\hour by60 \global\advance\minute
by-\hour}\edef\standardtime{{\ifnum\hour<12
\global\amorpm={am}%
        \else\global\amorpm={pm}\advance\hour by-12 \fi
        \ifnum\hour=0 \hour=12 \fi
        \number\hour:\ifnum\minute<10
0\fi\number\minute\the\amorpm}}
\edef\militarytime{\number\hour:\ifnum\minute<10
0\fi\number\minute}

\def\draftlabel#1{{\@bsphack\if@filesw {\let\thepage\relax
   \xdef\@gtempa{\write\@auxout{\string
      \newlabel{#1}{{\@currentlabel}{\thepage}}}}}\@gtempa
   \if@nobreak \ifvmode\nobreak\fi\fi\fi\@esphack}
        \gdef\@eqnlabel{#1}}
\def\@eqnlabel{}
\def\@vacuum{}
\def\draftmarginnote#1{\marginpar{\raggedright\scriptsize\tt#1}}
\def\draft{\oddsidemargin -.5truein
        \def\@oddfoot{\sl preliminary draft \hfil
        \rm\thepage\hfil\sl\today\quad\militarytime}
        \let\@evenfoot\@oddfoot \overfullrule 3pt
        \let\label=\draftlabel
        \let\marginnote=\draftmarginnote

\def\@eqnnum{(\theequation)\rlap{\kern\marginparsep\tt\@eqnlabel}%
\global\let\@eqnlabel\@vacuum}  }

\def\numberbysection{\@addtoreset{equation}{section}
        \def\theequation{\thesection.\arabic{equation}}}

\def\underline#1{\relax\ifmmode\@@underline#1\else
 $\@@underline{\hbox{#1}}$\relax\fi}

\catcode`@=12
\relax

\numberbysection

\advance \topmargin by -\headheight
\advance \topmargin by -\headsep

\evensidemargin \oddsidemargin
\def\br{\begin{eqnarray}}
\def\er{\end{eqnarray}}
\def\be{\begin{equation}}
\def\ee{\end{equation}}

\def\({\left(}
\def\){\right)}

\relax

\def\pa{\partial}

\def\tp0{\Theta_{+}^{(0)}}
\def\tm0{\Theta_{-}^{(0)}}

\def\f#1#2#3 {f^{#1#2}_{#3}}

\def\win1{{\sf w_{1+\infty}}}

\def\Win1{{\sf W_{1+\infty}}}

\def\rlx{\relax\leavevmode}
\def\inbar{\vrule height1.5ex width.4pt depth0pt}
\def\IZ{\rlx\hbox{\sf Z\kern-.4em Z}}
\def\IR{\rlx\hbox{\rm I\kern-.18em R}}
\def\IC{\rlx\hbox{\,$\inbar\kern-.3em{\rm C}$}}
\def\IN{\rlx\hbox{\rm I\kern-.18em N}}
\def\IO{\rlx\hbox{\,$\inbar\kern-.3em{\rm O}$}}
\def\IP{\rlx\hbox{\rm I\kern-.18em P}}
\def\IQ{\rlx\hbox{\,$\inbar\kern-.3em{\rm Q}$}}
\def\IF{\rlx\hbox{\rm I\kern-.18em F}}
\def\IG{\rlx\hbox{\,$\inbar\kern-.3em{\rm G}$}}
\def\IH{\rlx\hbox{\rm I\kern-.18em H}}
\def\II{\rlx\hbox{\rm I\kern-.18em I}}
\def\IK{\rlx\hbox{\rm I\kern-.18em K}}
\def\IL{\rlx\hbox{\rm I\kern-.18em L}}
\def\one{\hbox{{1}\kern-.25em\hbox{l}}}
\def\0#1{\relax\ifmmode\mathaccent"7017{#1}%
B        \else\accent23#1\relax\fi}

\begin{document}

\begin{titlepage}

\vspace{.2in}
\begin{center}
{\large\bf Modified non-linear Schr\"odinger models,  ${\cal C}{\cal P}_s{\cal T}_d$ symmetry, dark solitons and infinite towers of anomalous charges }\cal 
\end{center}

\vspace{.2in}

\begin{center}

H. Blas$^{(a)}$,  M. Cerna Magui\~na$^{(b)}$ and  L.F. dos Santos$^{(c)}$

\par \vskip .2in \noindent

$^{(a)}$Instituto de F\'{\i}sica\\
Universidade Federal de Mato Grosso\\
Av. Fernando Correa, $N^{0}$ \, 2367\\
Bairro Boa Esperan\c ca, Cep 78060-900, Cuiab\'a - MT - Brazil. \\
$^{(b)}$ Departamento de Matem\'atica\\
Universidad Nacional Santiago Ant\'unez de Mayolo\\
Campus Shancay\'an, Av. Centenario 200, Huaraz - Per\'u\\
$^{(c)}$ Centro Federado de Educa\c c\~ao Tecnologica-CEFET-RJ\\
Campus Angra dos Reis, Rua do Areal, 522, Angra dos Reis- RJ -Brazil

\normalsize
\end{center}

\vspace{.3in}

\begin{abstract}
\vspace{.3in}

Some modified (defocusing) non-linear Schr\"odinger models (MNLS)  possess infinite towers of anomalous conservation laws with asymptotically conserved charges. The so-called anomalies of the quasi-conservation laws vanish upon space-time integration  for a special ${\cal C}{\cal P}_s{\cal T}_d$ symmetric field configurations. We verify numerically the degree of modifications of the charges around the dark-soliton interaction regions by computing numerically some representative anomalies related to lowest order quasi-conservation laws of the non-integrable cubic-quintic NLS model as a modified (defocusing) NLS model. This modification depends on the parameter $\epsilon$, such that the standard NLS is recovered for $\epsilon=0$. Here we present the  numerical simulations for small values of $|\epsilon|$, and show that the collision of two dark solitons are elastic. The NLS-type equations are quite ubiquitous in several areas of non-linear science.  
\end{abstract}

\end{titlepage}

\section{Introduction}

Some non-linear field theory models with important physical applications and solitary wave solutions are not integrable. Recently, some deformations of integrable models, such as sine-Gordon, Korteweg-de Vries and non-linear Schr\"odinger models \cite{jhep1}\cite{blas1}\cite{blas3}\cite{blas5}, which exhibit soliton-type properties, have been put forward. Quasi-integrability properties of the deformations of the integrable models have recently been examined in the frameworks of the anomalous zero-curvature formulations \cite{jhep1}\cite{blas2}\cite{blas5} and the deformations of the Riccati-type pseudo-potential approach \cite{blas1}\cite{blas3}\cite{blas4}.  Recently, it has been considered the properties of the modified (focusing) non-linear Schr\"odinger model with bright solitons \cite{blas4}. Here we tackle the problem of constructing, analytically and numerically, new towers of anomalous charges for the  modified (defocusing) non-linear Schr\"odinger model with dark solitons; so extending the results of \cite{blas2} by providing novel infinite towers of quasi-conservation laws. The both type of models (focusing and defocusing) differ in the relevant signs ($+/-$) of their coupling constants and the boundary conditions (b.c.) imposed on their soliton solutions. So, in the focusing (defocusing) case one has bright (dark) solitons with vanishing (non-vanishing) b.c.'s.
\section{Quasi-conservation laws and anomalous charges}
Consider the modified non-linear Schr\"odinger models (MNLS) 
\br
\label{mnls}
 i \partial_{t} \psi(x,t) + \partial^2_{x} \psi(x,t) -  [\frac{\delta V( |\psi|^2)}{\delta |\psi|^2} ] \psi(x,t) =  0,
\er 
where $\psi \in C$ and $V: R_{+} \rightarrow R$ is the deformed potential.

Let us consider a special space-time reflection around a fixed point $(x_{\Delta},t_{\Delta})$ as a symmetry of soliton-type solutions of the model 
\br
\label{par1}
\widetilde{{\cal P}}:  (\widetilde{x},\widetilde{t}) \rightarrow (-\widetilde{x},-\widetilde{t});\,\,\,\,\,\,\,\,\widetilde{x} = x - x_{\Delta},\,\,\widetilde{t} = t- t_{\Delta}. 
\er 
The transformation $\widetilde{{\cal P}}$ defines a shifted parity ${\cal P}_{s}$ for the spatial variable $x$  and a delayed time reversal ${\cal T}_d$ for the time variable $t$. It is assumed that the  $\psi$ solution of the deformed NLS model (\ref{mnls})  possesses the following property under the transformation (\ref{par1})
\br
\label{par2}
\widetilde{{\cal P}} \equiv {\cal P}_{s}{\cal T}_{d},\,\,\,\,\,\,\,\,\widetilde{{\cal P}}(\psi) = e^{i \delta} \bar{\psi},\,\,\,\widetilde{{\cal P}}(\bar{\psi}) = e^{-i \delta} \psi,\,\,\,\,\bar{\psi} \equiv \psi^{\star},\,\,\,\,\delta = \mbox{constant.}
\er

In \cite{blas4} it has been provided a method to construct an infinite number of towers of quasi-conservation laws.  Here we consider the lowest order and the first three towers of quasi-conservation laws and discuss them in the context of the defocusing NLS with dark soliton solutions.  

{\sl The first order charge} and its generalization becomes \footnote{Formally, one can assume $\deg{(\psi^{\pm 1})} = \deg{(\bar{\psi}^{\pm1})}=  \pm 1, \deg{(\pa_x)}=\frac{1}{2}\deg{(\pa_t)}=1$.}
\br
\label{q1d}
\frac{d  }{dt}  Q_{1}(t) &=&  \int \, dx\, \hat{\alpha}_1,\,\,\,\,\,\,\,\,\,\hat{\alpha}_1 \equiv  2 F^{(1)}(I)\, \pa_x [\pa_x \bar{\psi} \pa_x \psi]\\
Q_{1} (t) &=& \int_{-\infty}^{\infty} \, dx\, [i F(I) \frac{\bar{\psi}\pa_x \psi - \psi \pa_x \bar{\psi}}{\bar{\psi}\psi}],\,\,\,\,\,\,I \equiv |\psi|^2,
\er
where $F^{(n)}(I) \equiv \frac{d^n}{dI^n} F(I)$. For $F=1$ one has $ \hat{\alpha}_1=0$ in  (\ref{q1d}) and the relevant charge $Q^{top}_1$ turns out to be the topological charge of the dark soliton's phase.
For arbitrary $F$ and the special solutions satisfying the parity property (\ref{par1})-(\ref{par2}) one has
\br
\int^{\widetilde{t}}_{-\widetilde{t}} dt \, \int^{\widetilde{x}}_{-\widetilde{x}} dx \,  \hat{\alpha}_1=0,\,\,\,\, \mbox{for} \,\,\,\widetilde{t} \rightarrow \infty,\,\,\,\widetilde{x} \rightarrow  \infty.
\er
Therefore, integrating in $t$ on the b.h.s.'s of (\ref{q1d}) one can get
\br
Q_{1} (\widetilde{t}) =Q_{1} (- \widetilde{t}),\,\,\,\, \widetilde{t} \rightarrow \infty.
\er
 The special function $F (I)=e^{-I^2}$ has been used in \cite{turbu} to study the first integrals in the study of soliton-gas and integrable turbulence.

A tower of  infinite number of quasi-conservation laws can be constructed on top of a given lowest order exact (quasi-)conservation law.  Next, we will present the first few of them. 
 
{\sl First tower}

One can construct a tower of quasi-conserved charges on top of the exact conserved charge ${\cal Q}_{1} = \frac{1}{2}\int\, dx \, (\bar{\psi}  \psi)$. So, one has
\br
\label{q2d}
\frac{d}{dt} {\cal Q}_{n} &=&\int dx\, \hat{\beta}_n;\,\,\,\,\,\,\,\hat{\beta}_n \equiv -\frac{1}{2n} \pa_x [(\bar{\psi} \psi)^{n-1}] i (\bar{\psi} \pa_x \psi - \psi \pa_x \bar{\psi} ),\,\,\,n=2,3,...\\
{\cal Q}_{n} &=& \frac{1}{2n}\int\, dx \, (\bar{\psi}  \psi)^n .\label{beta2d}
\er

For the field $\psi$ satisfying (\ref{par1})-(\ref{par2}) the anomaly density  $\hat{\beta}_n$ possesses an odd parity for any $n$. Therefore, one must have the vanishing of the space-time integral of the anomaly $\hat{\beta}_n$ and then, the asymptotically conserved charges satisfy 
\br
{\cal Q}_{n}(\widetilde{t}) = {\cal Q}_{n} (-\widetilde{t}),\,\,\,\, \widetilde{t} \rightarrow \infty,\,\,\,\,\,n=2,3,...
\er

{\sl Second tower}

The next tower of quasi-conserved charges is constructed on top of the exact conserved charge $\widetilde{Q}_{1} (t) = i  \int\, dx \, (\bar{\psi}\pa_x \psi - \psi \pa_x \bar{\psi}) $. So, one has

\br
\label{q3d}
\frac{d}{dt} \widetilde{Q}_{n}  = \int \, dx\, \hat{\gamma}_n ;\,\,\,\,\,\hat{\gamma}_n  &\equiv & - \frac{(i)^n}{n} \pa_x [(\bar{\psi}\pa_x \psi - \psi \pa_x \bar{\psi}) ^{n-1}]\times \\
&& \Big[ 2 \pa_x \psi \pa_x \bar{\psi} -\psi \pa^2_x \bar{\psi} - \bar{\psi} \pa_x^2 \psi + 2  V^{(1)}  |\psi|^2 - 2 V\Big],\nonumber\\
\widetilde{Q}_{n} (t) &=& \int\, dx \,\frac{(i)^n}{n} (\bar{\psi}\pa_x \psi - \psi \pa_x \bar{\psi}) ^n,\,\,\,\,\,n=2,3,....
\er
Similarly, one has  the asymptotically conserved charges
\br
\widetilde{Q}_{n}(\widetilde{t}) &=& \widetilde{Q}_{n}(-\widetilde{t}),\,\,\,\, \widetilde{t} \rightarrow \infty,\,\,\,\,n=2,3,...
\er 

{\sl Third tower}

The next tower of quasi-conserved charges is constructed on top of a fourth order one, which is itself quasi-conserved. One has
\br
\label{q4kd} 
\frac{d}{dt} K_n &=& \int dx\, \hat{\delta}_n,\,\,\,\,\,\,\,\,\, n=1,2,...
\\
\label{q4d1}
K_n &=& \int \, dx \, [\pa_{x} \bar{\psi} \pa_x{\psi}]^{n}.
\er
 The general form of the anomalies $ \hat{\delta}_n$ are provided in \cite{blas4}. Below we will consider the case $n=1$
\br
\label{an4d}
 \hat{\delta}_1&\equiv & i [(\bar{\psi} \pa_x \psi)^2-(\psi \pa_x\bar{\psi})^2] V^{(2)}(I),\,\,\,\,\,V^{(2)} \equiv \frac{d^2}{dI^2} V(I) .
\er

Regarding the standard (defocusing) NLS model one can argue that  all the anomalies will vanish upon integration in space-time provided that  the N-dark solitons satisfy  (\ref{par1})-(\ref{par2}). Consequently, their associated  charges will be  asymptotically conserved even for the standard NLS model. In fact, these type of solutions have been constructed in \cite{blas4} for N-bright solitons. We already have those results for standard NLS dark solitons and they will appear elsewhere. So, those examples show an analytical, and not only numerical, demonstration of the vanishing of the space-time integrals of the anomalies associated to the infinite towers of infinitely many quasi-conservation laws in soliton theory.  

\section{Numerical simulations}

We consider the non-integrable cubic-quintic NLS (CQNLS) model 
\begin{eqnarray}   
i \frac{\partial \psi(x,t) }{\partial t}  +  \frac{\partial^2 \psi(x,t) }{\partial x^2}  -\left( 2\eta |\psi(x,t)|^2 - \frac{\epsilon}{2} |\psi(x,t)|^4\right) \psi(x,t) =0,\
\label{cqnls2}
\end{eqnarray}
where $\eta > 0,\,\,\epsilon \in R$. The  model (\ref{cqnls2})  possesses a solitary wave solution  of the form $\psi(x,t) = \Phi(z) \mbox{exp}[i\Theta(z)+ i w t],\,z= x- v t$ (see \cite{blas2})
\br
\label{dark22}
\Phi^2_{\pm}(z) &=& \frac{\xi_{1} + r \xi_{2} \tanh^2{[k^{\pm} (z-z_0)]}}{ 1 + r \tanh^2{[k^{\pm} (z-z_0)]}}\\
\label{dark221}
\Theta_{\pm}(z) &=& \mp \arctan{\Big[ \sqrt{r \frac{\xi_{2}}{\xi_{1}}}  \, \,\tanh{[k^{\pm} (z-z_0)]}\Big]}
\er
where
\br
r & \equiv & \frac{|\psi_0|^2-\xi_{1}}{\xi_{2}-|\psi_0|^2},\,\,\,\,\,\,k^{\pm} \equiv \sqrt{\frac{|\epsilon|}{6}} \sqrt{(\pm ) (\xi_{2} - |\psi_0|^2) (|\psi_0|^2-\xi_{1})}\label{s22},\\
\xi_{1} &=& \frac{B - \sqrt{B^2- 6 v^2 \epsilon}}{2 \epsilon},
\xi_{2} = \frac{B + \sqrt{B^2- 6 v^2 \epsilon}}{2 \epsilon},\,\,\,\,\,\,
B \equiv 6 \eta -2 \epsilon |\psi_0|^2.
\er
The notations $\Phi_{\pm}$ and $\Theta_{\pm}$ correspond to $\epsilon>0$ and $\epsilon<0$, respectively.  So, we will take two one-dark solitary waves located some distance apart as the initial condition for our numerical simulations of two-dark soliton collisions. Below, we numerically compute the space and space-time integrals of the anomaly densities $\hat{\alpha}_1$,  $\hat{\beta}_2$,  $\hat{\gamma}_2$ and  $\hat{\delta}_1$, appearing in (\ref{q1d}), (\ref{q2d}), (\ref{q3d}) and (\ref{an4d}), respectively, for two type of two-soliton collisions of the CQNLS model (\ref{cqnls2}).

\begin{figure}
\centering
\includegraphics[width=1cm,scale=0.5, angle=0, height=4.31cm]{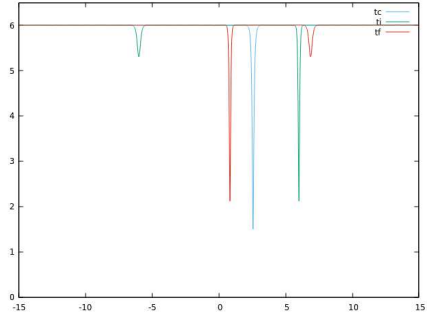} 
\caption {(color online) Collision of two dark solitons of the CQNLS model (\ref{cqnls2})  for $\epsilon= +0.005,\,|\psi_0|=6,\,\eta = 2.5$. The initial solitons ($t_i$ =green line) travel with velocities $v_1 \approx  - 4.7 \sqrt{2}$ (right soliton) and $v_2 \approx  13 \sqrt{2}$ (left soliton), respectively. They  completely overlap ($t_c$= blue line) in their closest approximation and then transmit to each other. The dark solitons after collision are plotted as a red line ($t_f$).}
\label{fig:Fig1}
\end{figure}
\begin{figure}
\centering
\includegraphics[width=8cm, scale=0.5, angle=0, height=4.8cm]{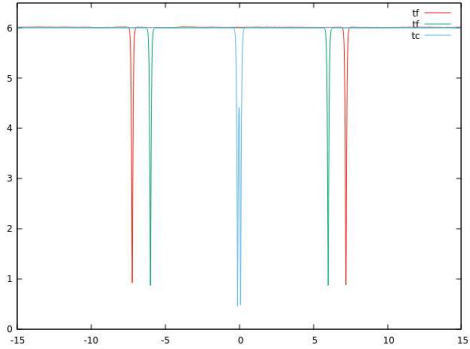} 
\caption {(color online) Reflection of two dark solitons of the cubic-quintic NLS model (\ref{cqnls2})  plotted for $\epsilon=-0.01,\,|\psi_0|=6,\,\eta = 2.5$. The initial solitons ($t_i$ =green line) travel in opposite direction with velocity $|v| \approx 1.97 \sqrt{2}$. They  partially overlap ($t_c$= blue line) in their closest approximation and then reflect to each other. The dark solitons after collision are plotted as a red line ($t_f$).}
\label{fig:Fig6}
\end{figure}

\subsection{Transmission of two-dark solitons and anomalous charges}

We numerically compute the space and space-time integrals of the anomaly densities $\hat{\alpha}_1$,  $\hat{\beta}_2$,  $\hat{\gamma}_2$ and  $\hat{\delta}_1$, as shown in Figs. 3, 4, 5 and 6, respectively, for the transmission of two-dark solitons of the CQNLS model (\ref{cqnls2}) as plotted in Fig. 1.
\begin{figure}
\centering
\epsfig{file=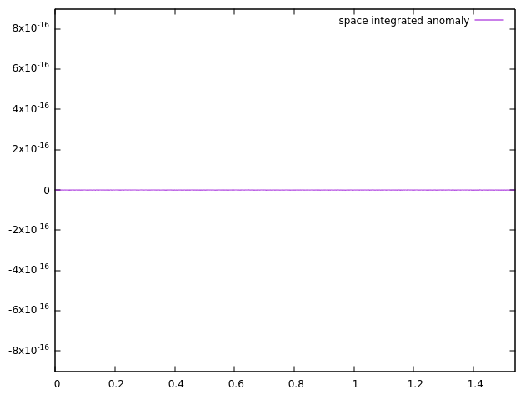, scale=1.11}
\epsfig{file=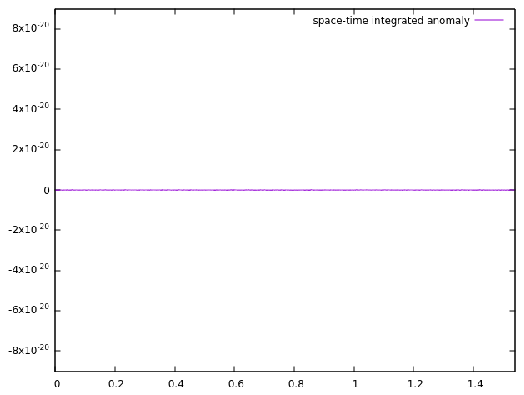, scale=1.11}
\caption {(color online)  The left Fig. shows the plot $\int_{-\widetilde{x}}^{+\widetilde{x}} \hat{\alpha}_1 dx\,\, vs\,\, t$ and the right one shows the plot $\int_{-\widetilde{t}}^{+\widetilde{t}}  dt \int_{-\widetilde{x}}^{+\widetilde{x}}  dx\, \hat{\alpha}_1\,\, vs\,\, t$ for the anomaly $\hat{\alpha}_1$ in  (\ref{q1d}) and for  the 2-soliton collision of Fig. 1.}
\label{fig:Fig2}
\end{figure}

\begin{figure}
\centering
\epsfig{file=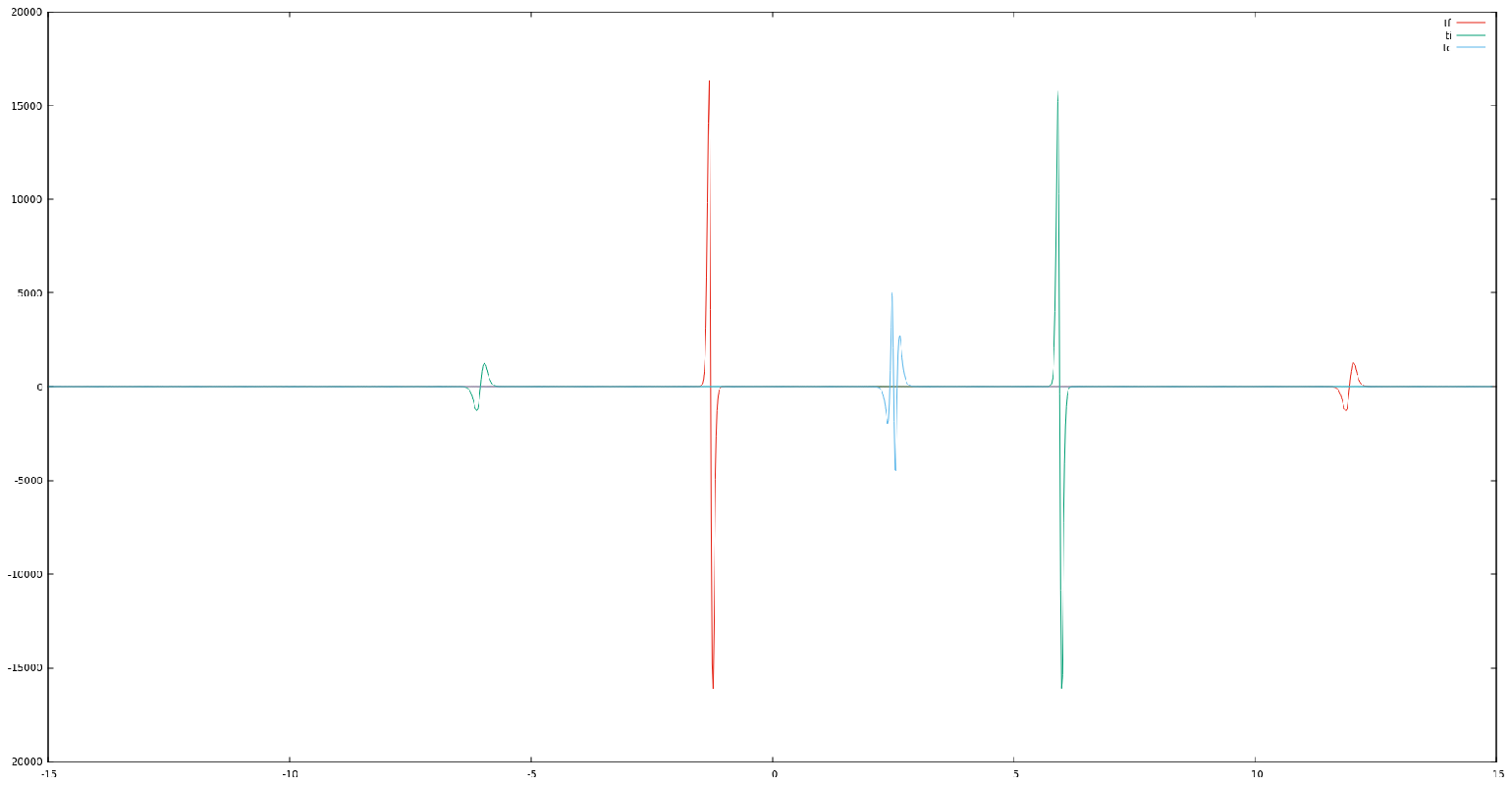, scale=0.7}
\epsfig{file=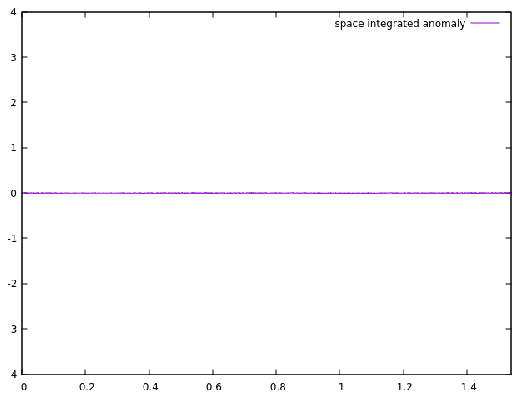, scale=1.1}
\epsfig{file=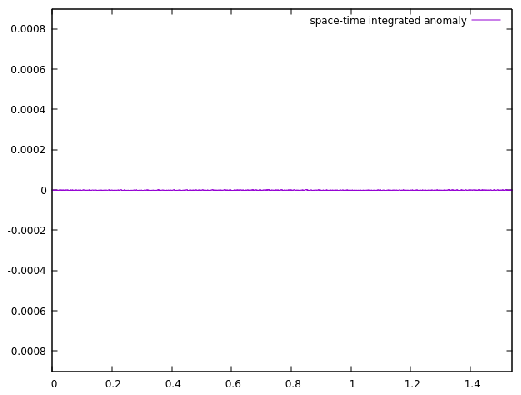, scale=1.1}
\caption {(color online) Top Fig. shows the  profile at initial (green), collision (blue) and final (red) times of the anomaly density $\hat{\beta}_2 $ in (\ref{q2d}) for the 2-soliton collision of Fig. 1. In the bottom the left Fig.  shows the plot $\int_{-\widetilde{x}}^{+\widetilde{x}} \hat{\beta}_2 dx\,\, vs\,\, t$ and the right one shows the plot $\int_{-\widetilde{t}}^{+\widetilde{t}}  dt \int_{-\widetilde{x}}^{+\widetilde{x}}  dx\, \hat{\beta}_2\,\, vs\,\, t$.}
\label{fig:Fig3}
\end{figure}

\begin{figure}
\centering
\epsfig{file=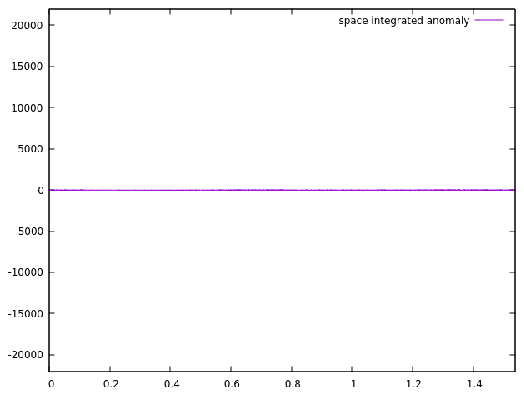, scale=1.1}
\epsfig{file=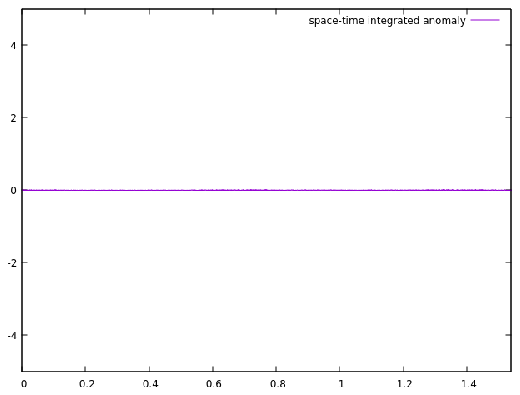, scale=1.1}
\caption {(color online) The left Fig.  shows the plot $\int_{-\widetilde{x}}^{+\widetilde{x}} \hat{\gamma}_2 dx\,\, vs\,\, t$ and the right one shows the plot $\int_{-\widetilde{t}}^{+\widetilde{t}}  dt \int_{-\widetilde{x}}^{+\widetilde{x}}  dx\, \hat{\gamma}_2\,\, vs\,\, t$ of the anomaly density $\hat{\gamma}_2 $ in (\ref{q3d}) for the 2-soliton collision of Fig. 1.}
\label{fig:Fig4}
\end{figure}

\begin{figure}
\centering
\epsfig{file=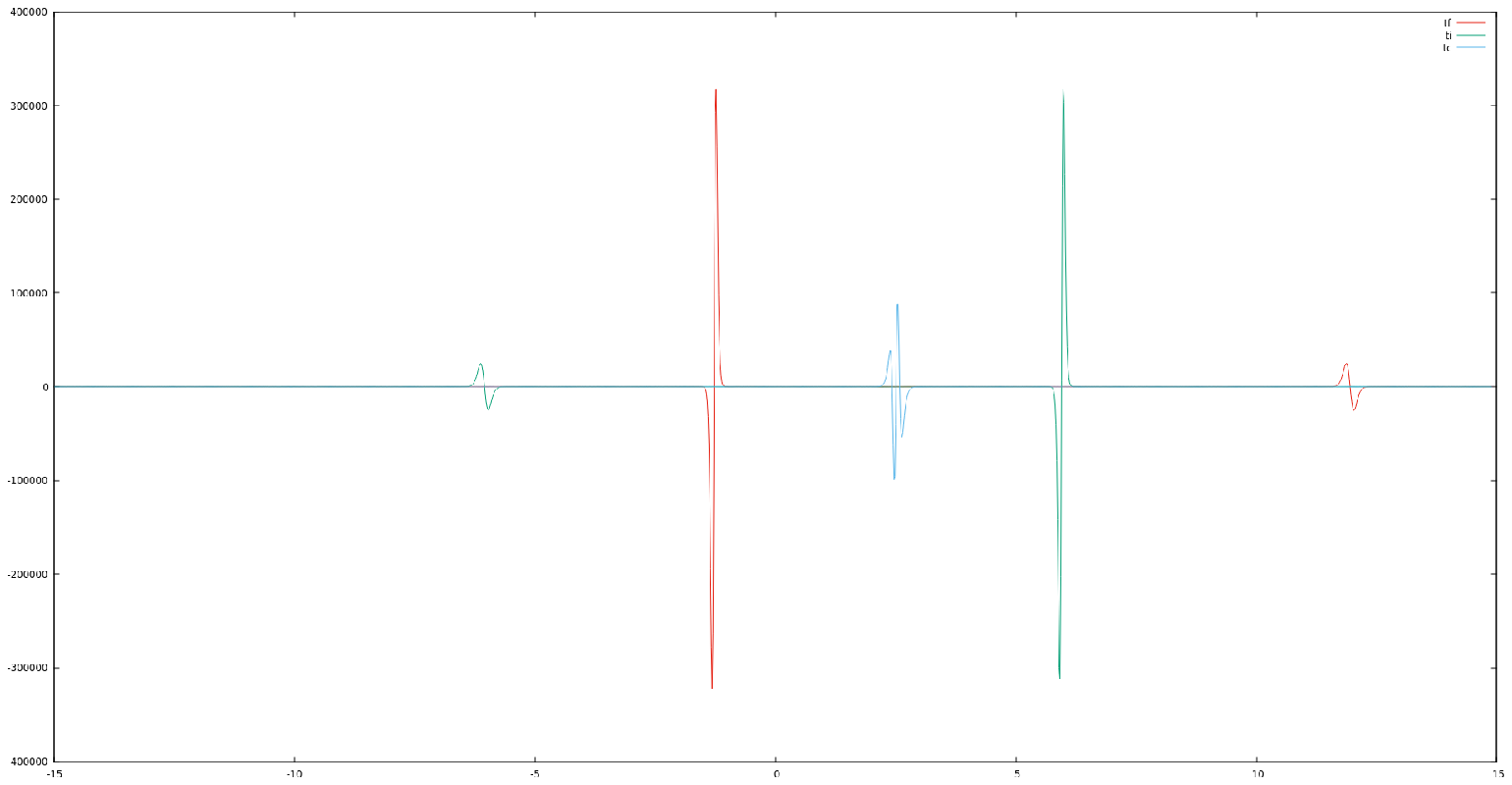, scale=0.7}
\epsfig{file=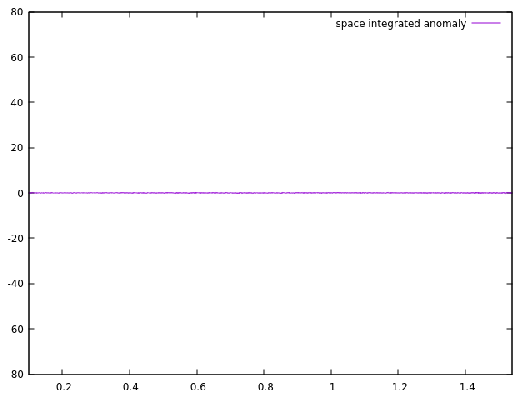, scale=1.1}
\epsfig{file=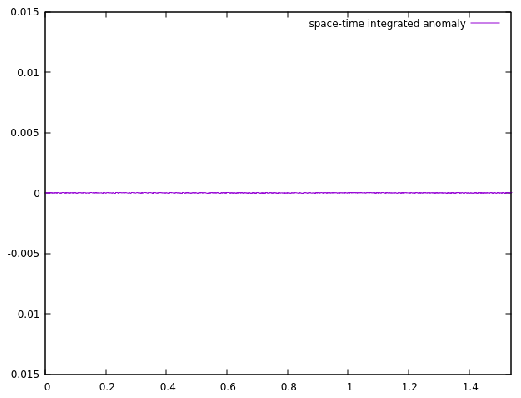, scale=1.1}
\caption {(color online) Top Fig. shows the  profile at initial (green), collision (blue) and final (red) times of the anomaly density $\hat{\delta}_1$ in  (\ref{an4d}) for the 2-soliton collision of Fig. 1. The bottom left shows the plot  $\int_{-\widetilde{x}}^{+\widetilde{x}} \hat{\delta}_1 dx\,\, vs\,\, t$ and the right one shows the plot $\int_{-\widetilde{t}}^{+\widetilde{t}}  dt \int_{-\widetilde{x}}^{+\widetilde{x}}  dx\, \hat{\delta}_1\,\, vs\,\, t$.}
\label{fig:Fig5}
\end{figure}
\newpage

\subsection{Reflection of two-dark solitons and anomalous charges}
 
Next, we numerically simulate the space and space-time integrals of the anomaly densities $\hat{\alpha}_1$,  $\hat{\beta}_2$,  $\hat{\gamma}_2$ and  $\hat{\delta}_1$, as shown in Figs. 7, 8, 9 and 10, respectively,  for the reflection of two-dark solitons of the CQNLS model (\ref{cqnls2}) as plotted in Fig. 2.

\begin{figure}
\centering
\epsfig{file=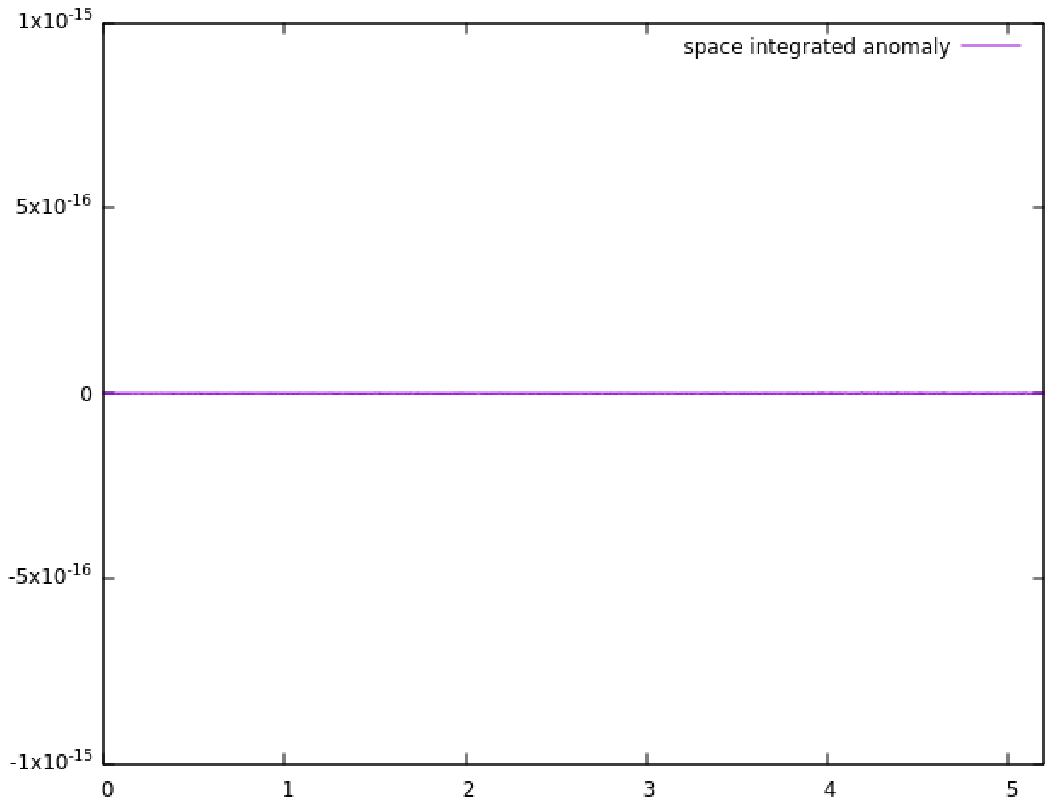, scale=0.5}
\epsfig{file=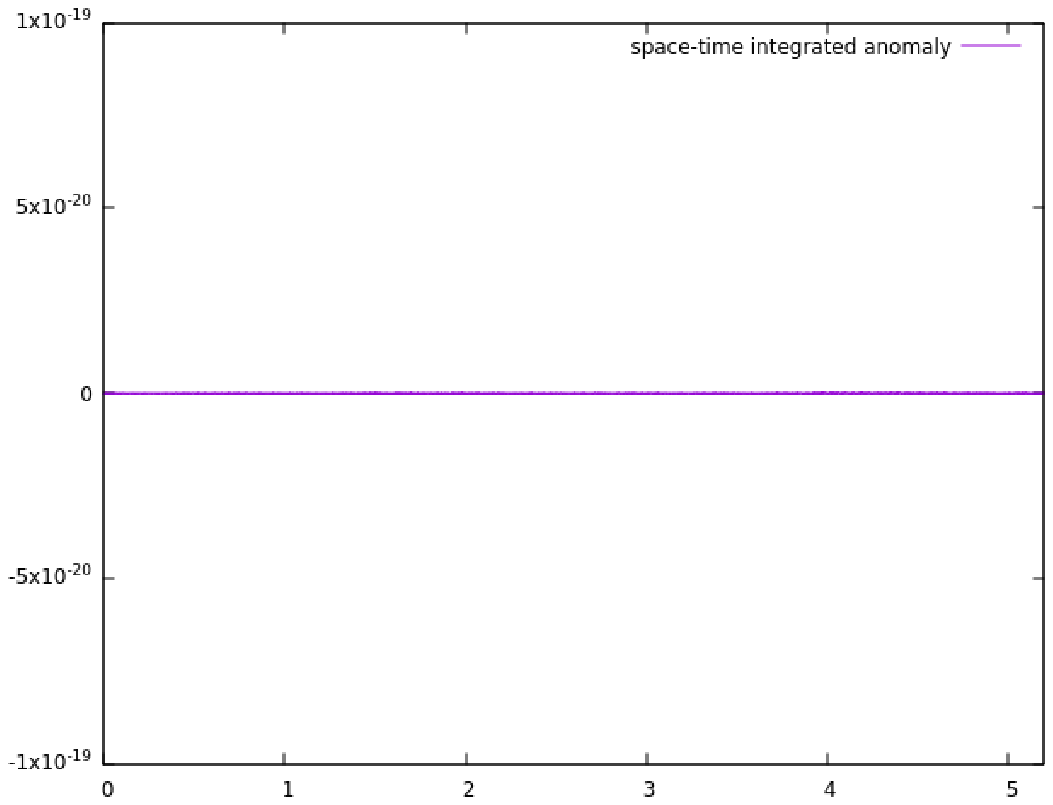, scale=0.5}
\caption {(color online) The left Fig. shows the plot $\int_{-\widetilde{x}}^{+\widetilde{x}} \hat{\alpha}_1 dx\,\, vs\,\, t$ and the right one shows the plot $\int_{-\widetilde{t}}^{+\widetilde{t}}  dt \int_{-\widetilde{x}}^{+\widetilde{x}}  dx\, \hat{\alpha}_1\,\, vs\,\, t$, for the anomaly $\hat{\alpha}_1$ in  (\ref{q1d}) computed for the soliton reflection in Fig. 2.}
\label{fig:Fig7}
\end{figure}

\begin{figure}
\centering
\epsfig{file=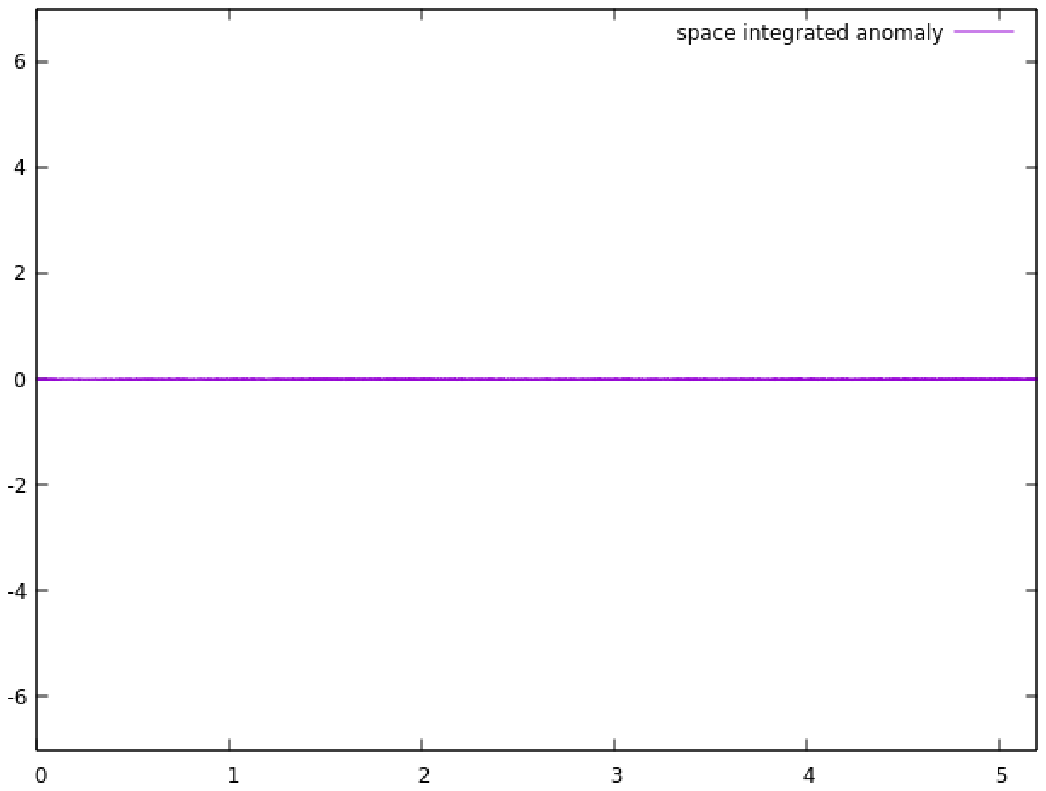, scale=0.52}
\epsfig{file=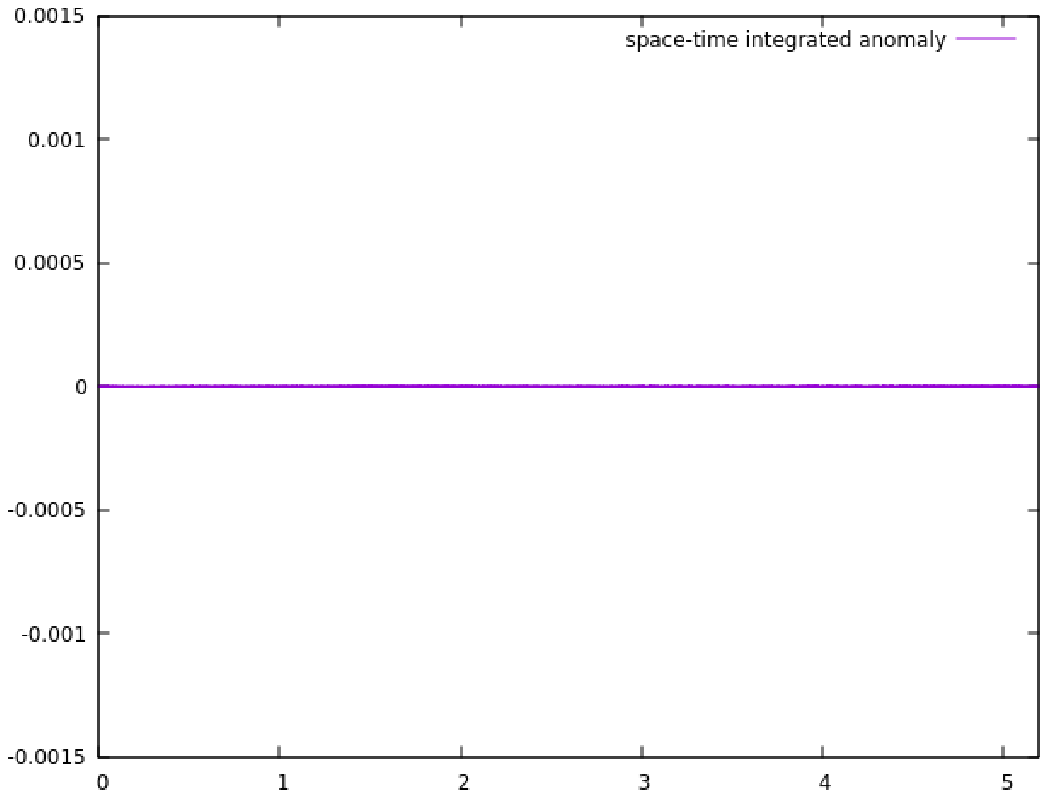, scale=0.52}
\caption {(color online) The left Fig.  shows the plot $\int_{-\widetilde{x}}^{+\widetilde{x}} \hat{\beta}_2 dx\,\, vs\,\, t$ and the right one shows the plot $\int_{-\widetilde{t}}^{+\widetilde{t}}  dt \int_{-\widetilde{x}}^{+\widetilde{x}}  dx\, \hat{\beta}_2\,\, vs\,\, t$ for the anomaly $\hat{\beta}_2$ in  (\ref{q2d}) computed for the soliton reflection in Fig. 2.}
\label{fig:Fig8}
\end{figure}

\begin{figure}
\centering
\epsfig{file=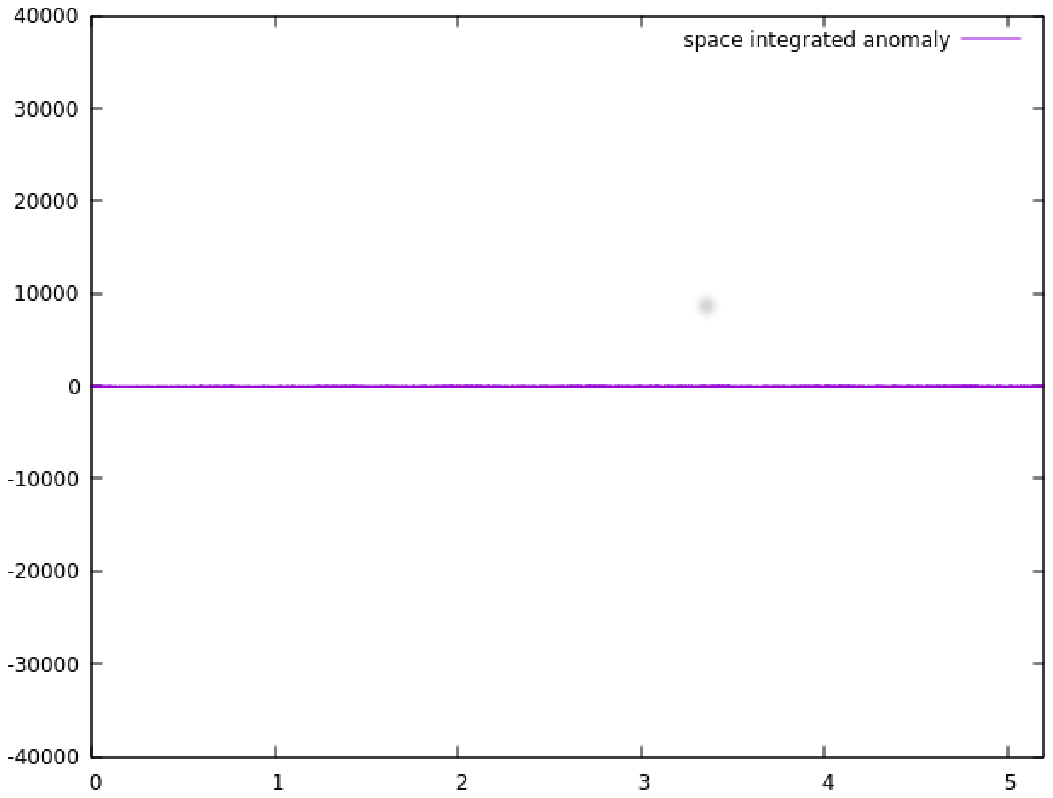, scale=0.51}
\epsfig{file=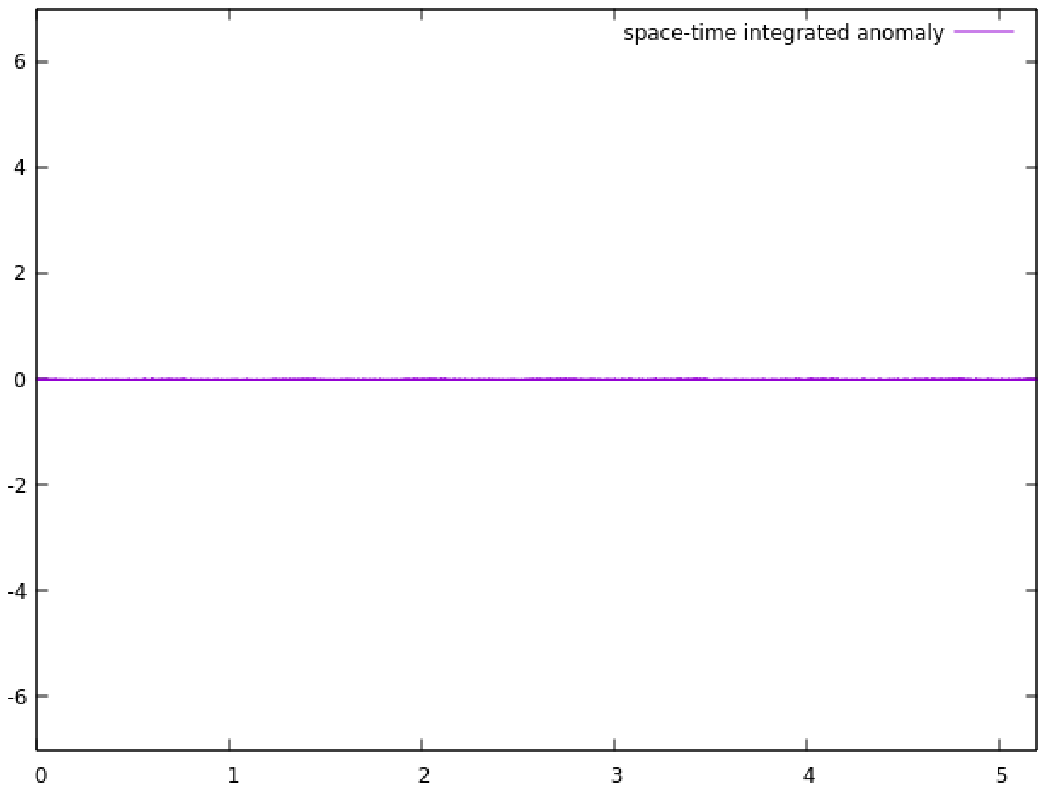, scale=0.51}
\caption {(color online) The left Fig.  shows the plot $\int_{-\widetilde{x}}^{+\widetilde{x}} \hat{\gamma}_2 dx\,\, vs\,\, t$ and the right one shows the plot $\int_{-\widetilde{t}}^{+\widetilde{t}}  dt \int_{-\widetilde{x}}^{+\widetilde{x}}  dx\, \hat{\gamma}_2\,\, vs\,\, t$ for the anomaly $\hat{\gamma}_2$ in   (\ref{q3d}) computed for the soliton reflection in Fig. 2.}
\label{fig:Fig9}
\end{figure}

\begin{figure}
\centering
\epsfig{file=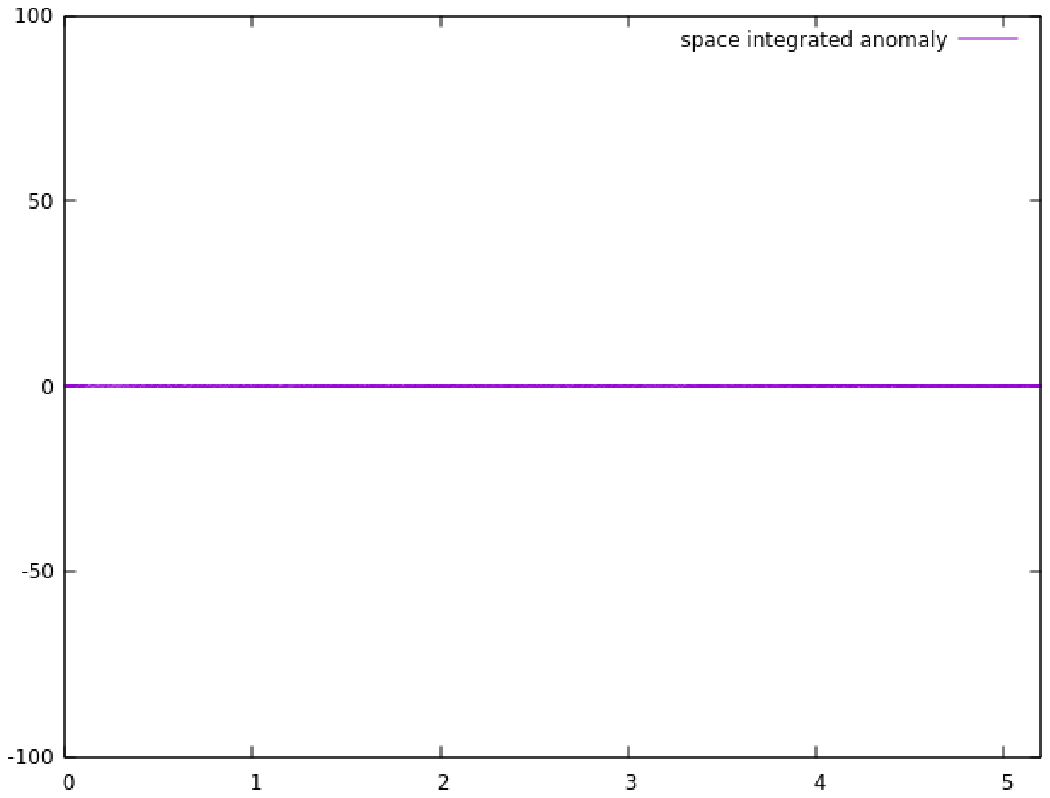, scale=0.51}
\epsfig{file=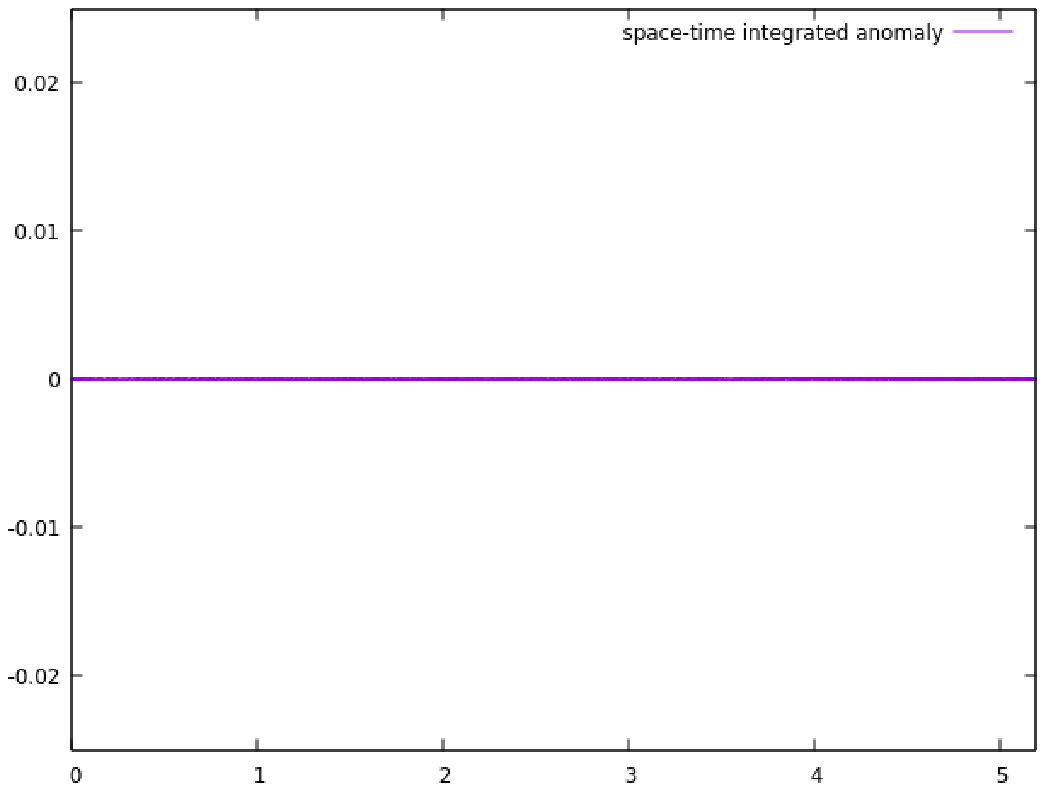, scale=0.51}
\caption {(color online) The left Fig.  shows the plot $\int_{-\widetilde{x}}^{+\widetilde{x}} \hat{\delta}_1 dx\,\, vs\,\, t$ and the right one shows the plot $\int_{-\widetilde{t}}^{+\widetilde{t}}  dt \int_{-\widetilde{x}}^{+\widetilde{x}}  dx\, \hat{\delta}_1\,\, vs\,\, t$ for the anomaly $\hat{\delta}_1$ in  (\ref{an4d}) computed for the soliton reflection in Fig. 2.}
\label{fig:Fig10}
\end{figure}

Some comments are in order here.  First, in our numerical simulations of the 2-dark soliton collisions  of the CQNLS model (\ref{cqnls2}) we have not observed appreciable emission of radiation during the collisions; so, it can be argued that the linear superposition of well separated two solitary waves of the CQNLS model  is an adequate initial condition. Second, we have shown  the vanishing of the space-time integrals of the anomaly densities $\hat{\alpha}_1$,  $\hat{\beta}_2$,  $\hat{\gamma}_2$ and  $\hat{\delta}_1$, appearing in  (\ref{q1d}), (\ref{q2d}), (\ref{q3d}) and (\ref{an4d}), respectively, within numerical accuracy. Third, we have performed extensive numerical simulations for a wide range of values in the parameter space; i.e. the deformation parameter $ |\epsilon| < 1$ and coupling constant $\eta \approx 2.5$, several amplitudes and relative velocities for 2-soliton collisions, obtaining the vanishing of those anomalies, within numerical accuracy. 

Sometimes the vanishing of the anomaly, within numerical accuracy, already happens for the space integration alone, e.g. as in the Figs. 3 and 7.  In fact, in the Fig. 7 one has $\int_{-\widetilde{x}}^{+\widetilde{x}}\, dx \, \hat{\alpha}_1 \approx 10^{-16}$. This fact can be explained by some symmetry considerations of the anomalies \cite{blas2} written in a new parametrization of the field $\psi$. So, let us write the anomaly density of eq. (\ref{q1d})  $\hat{\alpha}_1$ as 
\br
 \hat{\alpha}_1 = 2 F^{(1)}(I) \pa_x [ \frac{(\pa_x I)^2}{4 I} + \frac{1}{2} I (\pa_x \varphi)^2 ] , \,\,\,\, \psi \equiv \sqrt{I} e^{i\varphi/2}. 
\er    
Notice that this anomaly density is an odd function under the space reflection $x\rightarrow -x$, provided that $I\rightarrow I$  and  $\varphi \rightarrow \varphi$. In the Fig 2. for the collision  of two dark solitons one has the plot of the modulus $|\psi| = \sqrt{I}$ for three successive times which shows this type of symmetry for each time. 

On the other hand, the vanishing  $\int_{-\widetilde{x}}^{+\widetilde{x}} dx \hat{\alpha}_1 \approx 10^{-16}$ in Fig. 3 might happen for some  other reasons than the existence of some symmetry arguments as above, since this fact can not be visualized qualitatively in the collision of two dark solitons in Fig. 1.  In fact, the exact analytic 2-soliton solutions of the modified NLS model (\ref{cqnls2}) are not known, so it is not possible to show this type of symmetries for the explicit field configurations. However, in the case of the standard NLS model the analytic N-soliton solutions are available and the relevant space-time symmetries can be examined for the solutions and the various anomalies \cite{blas2}\cite{blas5}.     

The true understanding of the vanishing of the space-time integral of the anomalies and the relevance of them for the dynamics of the collision of the solitons of the  modified NLS model are under current investigations. The only explanation, so far,  for the vanishing  of the integrated anomalies, is the symmetry argument as presented above.  

Remarkably, infinite number of anomalies and the related quasi-conserved charges are also present in the standard  NLS model \cite{blas4}. So, an exact conserved charge of certain order can be constructed as a linear combination of some quasi-conserved charges of the same order, and when a linear combination of their related anomalies vanish, even before the space-time integration of them are performed.      

\section{Some conclusions and discussions}

Quasi-integrability properties of the CQNLS model have been examined by providing novel anomalous charges related to  infinite towers of  quasi-conservation laws. The anomaly densities exhibit odd parities under the  special space-time  symmetry (\ref{par1})-(\ref{par2}) of the field configurations.  
 
Through numerical simulations of $2-$dark soliton collisions we have checked the  quasi-conservation properties of the lowest order charges of the CQNLS model defined in (\ref{q1d}), (\ref{q2d}), (\ref{q3d}) and (\ref{q4kd}), respectively. So, we computed the space and space-time integrals of their associated anomaly densities $\hat{\alpha}_1$,  $\hat{\beta}_2$,  $\hat{\gamma}_2$ and  $\hat{\delta}_1$, for two types of two-soliton collisions of the CQNLS model (\ref{cqnls2}) as plotted in the Figs. 1 and 2, respectively.  In our numerical simulations presented in the Figs 3-6 and 7-10 we have observed that the space-time integrals of the set of anomalies $\hat{\alpha}_1$,  $\hat{\beta}_2$,  $\hat{\gamma}_2$ and  $\hat{\delta}_1$ vanish within numerical accuracy.  So, one can conclude that for 2-dark solitons the relevant charges are asymptotically conserved and their collisions are elastic within numerical accuracy, for a wide range of values of the set $\{\eta, \epsilon\}$ and a variety of amplitudes, velocities and relative initial phases. Since the modified NLS equations are quite ubiquitous, our results may find potential applications in several areas of non-linear science.

\section{Acknowledgments}

HB thanks FC-UNI (Lima-Per\'u) and FC-UNASAM (Huaraz-Per\'u) for hospitality during the initial stage of the work. MC thanks the Peruvian agency Concytec for partial financial support. LFdS thanks CEFET Celso Sukow da Fonseca-Rio de Janeiro-Brazil for kind support. The authors thank A. C. R. do Bonfim, H. F. Callisaya, C. A. Aguirre, J. P. R. Campos, R. Q. Bellido, J.M.J. Monsalve and A. Vilela for useful discussions.

%

\end{document}